\documentclass[aps,prd,twocolumn,nofootinbib,showpacs,showkeys]{revtex4}
\pdfoutput=1 

\usepackage{graphicx}
\usepackage{amsmath}
\usepackage{amssymb}
\usepackage{textcomp}
\usepackage{dcolumn}

\newcolumntype{k}{D{.}{.}{2,2}}

\newcommand{\e}[1]{$\times10^{#1}$}
\newcommand{\E}[1]{\times10^{#1}}
\newcommand{\nablabf}{\boldsymbol{\nabla}}
\newcommand{\new}[1]{#1}

\begin{document}

\title[Front propagation in thermoviscous fluids]{Analytical and numerical modeling of front propagation and interaction of fronts in nonlinear thermoviscous fluids including dissipation}

\author{Anders R.~\surname{Rasmussen}}\email{anders\_r\_r@yahoo.com}

\author{Mads P.~\surname{S\o rensen}}

\affiliation{Department of Mathematics, Technical University of Denmark, DK-2800
  Kongens Lyngby, Denmark}

\author{Yuri B.~\surname{Gaididei}}

\affiliation{Bogolyubov Institute for Theoretical Physics, 03680 Kiev, Ukraine}

\author{Peter L.~\surname{Christiansen}}

\affiliation{\new{Department of} Informatics and Mathematical Modelling and Department of Physics,
  Technical University of Denmark, DK-2800 Kongens Lyngby, Denmark}

\date{\today}

\begin{abstract}
A wave equation, that governs finite amplitude acoustic disturbances in a thermoviscous Newtonian fluid, and includes nonlinear terms up to second order, is proposed. In contrast to the model known as the Kuznetsov equation, the proposed nonlinear wave equation preserves the Hamiltonian structure of the fundamental fluid dynamical equations in the non-dissipative limit. An exact traveling front solution is obtained from a generalized traveling wave assumption. \new{This solution is, in an overall sense, equivalent to the Taylor shock solution of the Burgers equation. However, in contrast to the Burgers equation, the model equation considered here is capable to describe waves propagating in opposite directions.} Owing to the Hamiltonian structure of the proposed model equation, the front solution is in agreement with the classical Rankine-Hugoniot relations. The exact front solution propagates at supersonic speed with respect to the fluid ahead of it, and subsonic speed with respect to the fluid behind it, similarly to the fluid dynamical shock. Linear stability analysis reveals that the front is stable when the acoustic pressure belongs to a critical interval, and is otherwise unstable. These results are verified numerically. Studies of head-on colliding fronts demonstrate that the front propagation speed changes upon collision.
\end{abstract}



\maketitle

\section{Introduction}
The ``classical'' equation of nonlinear acoustics \cite{aanonsen}, the so-called
Kuznetsov equation \cite{kuznetsov}, governs finite amplitude acoustic
disturbances in a Newtonian, homogeneous, viscous, and heat conducting fluid.
The model equation and its paraxial approximation, the
Khokhlov-Zabolotskaya-Kuznetsov (KZK) equation \cite{zabolotskaya,kuznetsov}, are
occasionally encountered within studies related to nonlinear wave propagation.
See e.g.\ the recent works by Jordan \cite{jordan1} who presented the derivation
and analysis of an exact traveling wave solution to the one-dimensional
Kuznetsov equation, and by Jing and Cleveland \cite{jing} who described a
three-dimensional numerical code that solves a generalization of the KZK
equation, and the references cited in the introductory sections of those papers.
Other recent works based on the Kuznetsov equation include: analysis of energy
effects accompanying a strong sound disturbance \cite{wojcik1}, studies of
generation of higher harmonics and dissipation based on a 3D finite element
formulation \cite{hoffelner}, and studies of nonlinear wave motion in cylindrical
coordinates \cite{shermenev}. The derivations of the Kuznetsov
equation \cite{kuznetsov,makarov2,enflo} and related model
equations \cite{soderholm,naugolnykh,sapozhnikov} are based on the complete
system of the equations of fluid dynamics. It has been demonstrated that this
system of equations is of Hamiltonian structure in the non-dissipative
limit \cite{zakharov}. However, in the non-dissipative limit, the Kuznetsov equation
does not retain the Hamiltonian structure.

In this paper we propose a nonlinear wave equation, which, in the
non-dissipative limit, preserves the Hamiltonian structure of the fundamental
equations. Furthermore, we present the derivation and analysis of an exact
traveling front solution, which applies equally well to the proposed nonlinear
wave equation and the Kuznetsov equation. The derivation of the exact solution
is based on a \emph{generalized} traveling wave assumption, which leads to a
wider class of exact solutions compared to the one reported by
Jordan \cite{jordan1,jordan2}. Furthermore, the introduction of the generalized
assumption is necessary in order to interpret the results of numerical
simulations of head-on colliding fronts presented in this paper. In order to
relate our results to the classical literature, we demonstrate that the exact
front solution retains a number of properties of the fluid dynamical shock. The
paper is structured as follows: The proposed equation and its Hamiltonian
structure are discussed in Section~\ref{nlweq}. Section~\ref{etfs} contains the
derivation of our exact traveling front solution and analysis of its stability
properties. In Section~\ref{fsr} we demonstrate that the front is related the
classical shock.  Section~\ref{numstud} presents numerical investigations of the
front, while Section~\ref{concl} contains our conclusions.

\section{Nonlinear wave equations\label{nlweq}}
Equations governing finite amplitude acoustic disturbances in a Newtonian,
homogeneous, viscous and heat conducting fluid may be derived from the following
four equations of fluid dynamics: \emph{the equation of motion}
\begin{multline}\label{motion}
  \rho \left(\frac{\partial \mathbf{u}}{\partial t} + (\mathbf{u} \cdot
    \nablabf)\mathbf{u}\right) \\ = - \nablabf p + \eta \Delta \mathbf{u} +
  \left(\frac{\eta}{3}+\zeta\right) \nablabf(\nablabf \cdot \mathbf{u}),
\end{multline}
\emph{the equation of continuity}
\begin{align}\label{cont}
  \frac{\partial \rho}{\partial t} + \nablabf \cdot (\rho \mathbf{u}) = 0,
\end{align}
\emph{the heat transfer equation}
\begin{multline}\label{heat}
  \rho T \left(\frac{\partial s}{\partial t} + (\mathbf{u} \cdot
    \nablabf)s\right) = \frac{\eta}{2} \left(\frac{\partial u_i}{\partial
      x_j} +\frac{\partial u_j}{\partial x_i}
    -\frac{2}{3}\delta_{ij}\frac{\partial u_k}{\partial x_k}\right)^2 
\\ + \zeta(\nablabf \cdot \mathbf{u})^2 + \kappa \Delta T,
\end{multline}
and \emph{the equation of state}
\begin{align}\label{state}
  p=p(\rho,s).
\end{align}
Here $\mathbf{x} = \left(x,y,z\right)$ are the spatial (Cartesian) coordinates
and $t$ denotes time. $\mathbf{u} = \left(u,q,w\right)$ is the fluid particle
velocity, $\rho$ is the density of the medium, $p$, $s$, and $T$ are the
thermodynamic variables pressure, entropy and temperature, respectively. $\eta$
and $\zeta$ are the coefficients of shear and bulk viscosity, and $\kappa$ is
the heat conductivity coefficient. $\Delta$ is the Laplace operator.

To obtain a nonlinear wave equation all dependent variables except one are
eliminated from the system \eqref{motion}--\eqref{state}, resulting in a
nonlinear wave equation for that single
variable \cite{kuznetsov,makarov2,soderholm,enflo,naugolnykh,sapozhnikov}. The
deviations of $\rho$, $p$, $s$, and $T$ from their equilibrium values, $\rho_0$,
$p_0$, $s_0$, and $T_0$ are assumed to be small, as well as the fluid particle
velocity, $|\mathbf{u}|$. The heat conductivity coefficient $\kappa$ and the
viscosities $\eta$ and $\zeta$ are also treated as small quantities. In order to
obtain a second order approximation, all equations are written retaining terms
up to second order in the small quantities. It is assumed that the flow is
rotation free, $\nablabf \times \mathbf{u} = 0$, thus
\begin{align}\label{velpot}
  \mathbf{u} \equiv -\nablabf\psi,
\end{align}
where $\psi$ is the velocity potential. Furthermore, it has become customary to
use the following approximation for the equation of state \cite{makarov1}
\begin{multline}\label{approxState}
  p-p_0 = c_0^2 \left(\rho-\rho_0\right) + \frac{c_0^2}{\rho_0} \frac{B/A}{2}
  \left(\rho-\rho_0\right)^2 \\ + \left(\frac{\partial p}{\partial
      s}\right)_{\rho,s=s_0} \left(s-s_0\right),
\end{multline}
where $B/A$ is the fluid nonlinearity parameter \cite{beyer} and $c_0^2 \equiv
\left(\partial p/\partial \rho\right)_{s,\rho=\rho_0}$ is the small-signal sound
speed. Then, from Eqs.~\eqref{motion}--\eqref{approxState} we obtain the
following nonlinear wave equation
\begin{multline}\label{modkuz}
  \frac{\partial^2 \psi}{\partial t^2} - c_0^2 \Delta\psi = \frac{\partial
    \psi}{\partial t} \Delta\psi \\ + \frac{\partial}{\partial t}
  \left(b\Delta\psi + (\nablabf\psi)^2 + \frac{B/A-1}{2c_0^2}
    \left(\frac{\partial \psi}{\partial t}\right)^2\right),
\end{multline}
where $b$ is the diffusivity of sound \cite{hamilton1}
\begin{align}\label{diffu}
  b \equiv \frac{1}{\rho_0}\left\{\frac{4}{3}\eta +\zeta +
    \kappa\left(\frac{1}{C_V}-\frac{1}{C_p}\right)\right\},
\end{align}
and $ C_V$ and $C_p$ denote the heat capacities at constant volume and pressure,
respectively. Typical values of the physical parameters $c_0$, $B/A$, and $b$
are given in Table~\ref{parval}. In the first order approximation,
Eq.~\eqref{modkuz} reduces to
\begin{align}\label{lwe}
  \frac{\partial^2 \psi}{\partial t^2} = c_0^2 \Delta\psi.
\end{align}
Introducing Eq.~\eqref{lwe} in the first term on the right hand side of
Eq.~\eqref{modkuz}, the Kuznetsov equation \cite{kuznetsov}
\begin{align}\label{origkuz}
  \frac{\partial^2 \psi}{\partial t^2} - c_0^2 \Delta \psi =
  \frac{\partial}{\partial t}\left(b\Delta\psi+(\nablabf \psi)^2 +
    \frac{B/A}{2c_0^2}\left(\frac{\partial \psi}{\partial t}\right)^2\right),
\end{align}
is obtained.

\begin{table}
\caption{Values of $c_0$, $B/A$, and $b$ for three different substances. 
The values for $b$ are rough estimates obtained from Eq.~\eqref{diffu} 
neglecting the influence of bulk viscosity and thermal 
losses.}\label{parval}
\label{BAtab}
\begin{ruledtabular}
\begin{tabular}{l D{;}{~}{3,5} D{;}{~}{2,6}  D{;}{~}{8,6}}
Substance & \multicolumn{1}{c}{$c_0$ (m$\,$s$^{-1}$)} 
& \multicolumn{1}{c}{$B/A$} & \multicolumn{1}{c}{$b$ (m$^2\,$s$^{-1}$)} 
\\ \hline 
Water
& 1483;(20\text{\textcelsius})^{\text{a}}
& 5.0;(20\text{\textcelsius})^{\text{b}}
& 1.3\E{-6};(20\text{\textcelsius})^\text{c} \\
Air
& 343;(20\text{\textcelsius})^{\text{a}}
& 0.4;(20\text{\textcelsius})^{\text{b,d}}
& 21\E{-6};(27\text{\textcelsius})^\text{c} \\
Soft tissue
& 1540;\hspace{-0.5mm}^{\text{e}}
& 9.6;(37\text{\textcelsius})^{\text{b,f}}
& \multicolumn{1}{c}{\text{N/A}} \\
\end{tabular}
\end{ruledtabular}
\raggedright
{\footnotesize
$^{\text{a}}$ {Ref.~\onlinecite{crc}} \\
$^{\text{b}}$ {Ref.~\onlinecite{beyer}} \\
$^{\text{c}}$ {Values for $\rho_0$ and $\eta$ are obtained from
  Ref.~\onlinecite{crc}.} \\
$^{\text{d}}$ {Diatomic gas} \\
$^{\text{e}}$ {Ref.~\onlinecite{gent}} \\
$^{\text{f}}$ {Human breast fat}}
\end{table}

In absence of dissipation, i.e.\ $\eta=\zeta=0$, Eqs.~\eqref{motion} and \eqref{cont}
possess Hamiltonian structure \cite{zakharov}. This property is, however,
\emph{not} retained in Eq.~\eqref{origkuz} with $b = 0$, i.e.\ the non-dissipative 
limit of the Kuznetsov equation is not Hamiltonian. In contrast, Eq.~\eqref{modkuz}
\emph{does} retain the Hamiltonian structure in the non-dissipative limit.
Accordingly, the equation may be derived from the Lagrangian density
\begin{align}\label{lag}
\mathcal{L} = \frac{\left(\psi_t\right)^2}{2}
- c_0^2\frac{\left(\nablabf\psi\right)^2}{2}
- \frac{B/A-1}{6c_0^2}\left(\psi_t\right)^3 
- \frac{\psi_t\left(\nablabf \psi\right)^2}{2},
\end{align}
using the Euler-Lagrange equation\footnote{Letting $\eta=\zeta=0$, $\mathbf{u}=-\nablabf\psi$, and $p=\rho^\gamma/\gamma$ in Eqs.~(\ref{motion}--\ref{cont}), and \eqref{state}, respectively, one can derive the Lagrangian density
\begin{align*}
\mathcal{L}_{PEE} = \frac{c_0^4}{\gamma}\left(1 + \frac{\gamma-1}{c_0^2}\left(\psi_t - \frac{(\nablabf \psi)^2}{2}\right)\right)^{\dfrac{\gamma}{\gamma-1}},
\end{align*}
corresponding to the potential Euler equation (PEE) given in Ref.~\onlinecite{christov}. Expanding $\mathcal{L}_{PEE}$ to third order and letting $\gamma=B/A+1$ we obtain Eq.~\eqref{lag}.}. From the Legendre transformation \cite{goldstein} we obtain the corresponding Hamiltonian density as
\begin{align}\label{hamden}
  \mathcal{H} = c_0^2 \frac{\left(\nablabf \psi\right)^2}{2} +
  \frac{\left(\psi_t\right)^2}{2} - \frac{B/A-1}{3c_0^2}\left(\psi_t\right)^3,
\end{align}
which may be integrated to yield the total Hamiltonian
\begin{align}\label{tnrg}
  H = \int_{-\infty}^{+\infty}\int_{-\infty}^{+\infty}\int_{-\infty}^{+\infty}  \mathcal{H}\,dx\,dy\,dz.
\end{align}
\new{Taking the time derivative of $H$ in Eq.~\eqref{tnrg} with $\mathcal{H}$ replaced by Eq.~\eqref{hamden}, and using Eq.~\eqref{modkuz}, one can obtain a simple expression for $dH/dt$. Doing this in one spatial dimension we obtain after some calculations
\begin{align}\label{eqnrg}
  \frac{dH}{dt} = \left[c_0^2 \psi_t \psi_x + \left(\psi_t\right)^2
    \psi_x \right] _{-\infty}^{+\infty} - b \int_{-\infty}^{+\infty}
  \left(\psi_{xt}\right)^2 dx.
\end{align}
In Eq.~\eqref{eqnrg}, which is sometimes called the energy balance equation,} we observe that the first terms on the right hand side correspond to energy in- and output at the two boundaries, and that the last term accounts for energy dissipation inside the system.

In the remaining portion of this paper we shall limit the analysis to one-dimensional plane fields, in which case the proposed model equation \eqref{modkuz} reduces to
\begin{multline}\label{modkuz1D}
  \psi_{tt} - c_0^2 \psi_{xx} = \psi_t \psi_{xx} \\ + \frac{\partial}{\partial t}
  \left(b\psi_{xx} + \left(\psi_x\right)^2 + \frac{B/A-1}{2c_0^2}
    \left(\psi_t\right)^2\right),
\end{multline}
where subscripts indicate partial differentiation.

Finally, for later reference we give the second-order expressions for the
acoustic density, $\rho-\rho_0$, and acoustic pressure, $p-p_0$, in terms of the
velocity potential, $\psi$. From the equations of motion \eqref{motion} and
state \eqref{approxState}, subject to the basic assumptions of the derivation of
the two model equations \eqref{modkuz} and \eqref{origkuz}, we obtain
\begin{align}\label{sorho}
  \rho - \rho_0 &= \frac{\rho_0}{c_0^2} \left(\psi_t -
    \frac{\left(\psi_x\right)^2}{2} - \frac{B/A-1}{2c_0^2}\left(\psi_t\right)^2
    - b\psi_{xx}\right),
\end{align}
and
\begin{align}\label{sop}
  p - p_0 &= \rho_0 \left(\psi_t - \frac{\left(\psi_x\right)^2}{2} +
    \frac{1}{2c_0^2}\left(\psi_t\right)^2\right) - \left(\frac{4}{3}\eta
    +\zeta\right)\psi_{xx},
\end{align}
respectively. It should be noted that Eqs.~\eqref{sorho} and \eqref{sop} are
derived from the fundamental equations, thus, the expressions are not specific
to any of the two model equations \eqref{modkuz} and \eqref{origkuz}.

\section{Exact traveling front solution\label{etfs}}
Recently, a standard traveling wave approach was applied to the one-dimensional
approximation of the Kuznetsov equation \eqref{origkuz} to reveal an exact
traveling wave solution \cite{jordan1,jordan2}. In this section we extend the
standard approach by introducing a generalized traveling wave assumption and
analyze the stability properties of the solution.

\subsection{Generalized traveling wave analysis\label{gtwa}}
We introduce the following generalized traveling wave assumption
\begin{align}\label{mtwas}
\begin{split}
  \psi(x,t) &= \Psi(x-vt) - \lambda x + \sigma t \\ &\equiv \Psi(\xi) - \lambda
  x + \sigma t,
\end{split}
\end{align}
where $\lambda$ and $\sigma$ are arbitrary constants, $v$ denotes the wave
propagation velocity, and $\xi \equiv x-vt$ is a wave variable. \new{The inclusion of $- \lambda x + \sigma t$ in Eq.~\eqref{mtwas} leads to a wider class of exact solutions, compared to the one obtained from the assumption $\psi=\Psi(x-vt)$, which is the standard one. Furthermore, the introduction of the generalized assumption is necessary in order to interpret the results of numerical
simulations of head-on colliding fronts presented in Section~\ref{headsec}.}
Inserting Eq.~\eqref{mtwas} into the nonlinear wave equation \eqref{modkuz1D} we obtain the ordinary differential equation
\begin{multline}\label{theode}
  \left(v^2-c_0^2\right)\Psi'' = \left(-v\Psi'+\sigma\right)\Psi''  - 
  v\frac{d}{d\xi}\bigg\{b\Psi'' \\ + \left(\Psi' - \lambda\right)^2 +
    \frac{B/A-1}{2c_0^2}\left(-v\Psi'+\sigma\right)^2\bigg\},
\end{multline}
where prime denotes ordinary differentiation with respect to $\xi$. Integrating
once and introducing $\Phi \equiv -\Psi'$, Eq.~\eqref{theode} reduces to
\begin{multline}\label{twODEc}
  C = vb\Phi' - \left(\frac{3}{2} + \frac{B/A-1}{2c_0^2}v^2\right) v\Phi^2 + \\
  \left\{ \left( 1 - \frac{B/A - 1}{c_0^2}\sigma \right)v^2 -2\lambda v - c_0^2
    - \sigma \right\} \Phi,
\end{multline}
where $C$ is a constant of integration. Requiring that the solution satisfy
$\Phi' \to 0$ as $\xi \to \pm \infty$, and either
\begin{align}\label{bcphi}
  \Phi \to \left\{\begin{array}{ll} \theta , &\xi \to + \infty \\ 0, &\xi \to -
      \infty
\end{array} \right. 
\quad \text{or} \quad \Phi \to \left\{\begin{array}{ll} 0 , &\xi \to + \infty \\
    \theta, &\xi \to - \infty
\end{array} \right. ,
\end{align}
where $\theta$ is an arbitrary constant, lead us to $C=0$ and
\begin{multline}\label{cubic}
  \frac{B/A-1}{2c_0^2} \theta v^3 - \left(1 - \frac{B/A-1}{c_0^2} \sigma\right)
  v^2 \\ + \left(\frac{3}{2} \theta + 2\lambda\right) v + c_0^2 + \sigma = 0.
\end{multline}
In order to obtain our traveling wave solution, we separate the variables in
Eq.~\eqref{twODEc} subject to $C=0$, then, using Eq.~\eqref{cubic}, we find the
solution to be the traveling front
\begin{align}
  \Phi &= \frac{\theta}{2} \left\{1 -
    \tanh\left(\frac{2\left(\xi-x_0\right)}{l}\right)\right\},
\label{twsphi} \\ 
l &\equiv \frac{4b}{\left(\dfrac{B/A-1}{2c_0^2}v^2 + \dfrac{3}{2} \right)
  \theta},
\label{thick}
\end{align}
where $x_0$ is an integration constant, $|l|$ is the front thickness, and $0 <
\Phi < \theta$. Finally, using $\Phi=-\Psi'$ and inserting Eq.~\eqref{twsphi}
into Eq.~\eqref{mtwas} we obtain (apart from an arbitrary constant of
integration)
\begin{align}\label{twpsi}
\psi(x,t) = -\frac{\theta}{2} \left\{\xi - \frac{l}{2} \ln\left(\cosh\frac{2(\xi-x_0)}{l}\right)\right\} - \lambda x + \sigma t,
\end{align}
which is the exact solution for the velocity potential.

\new{Traveling $\tanh$ solutions, such as the front solution~\eqref{twsphi}, are often called Taylor shocks. The existence of an exact solution of this type to the classical Burgers equation is a well known result \cite{pierce}. However, the Burgers equation is restricted to wave propagation \emph{either} to the  left or to the right. The model equation considered in this paper does not suffer from this limitation, as shall be illustrated in Section~\ref{headsec}.}

\begin{figure}[h]
  \centering \includegraphics[width=8cm]{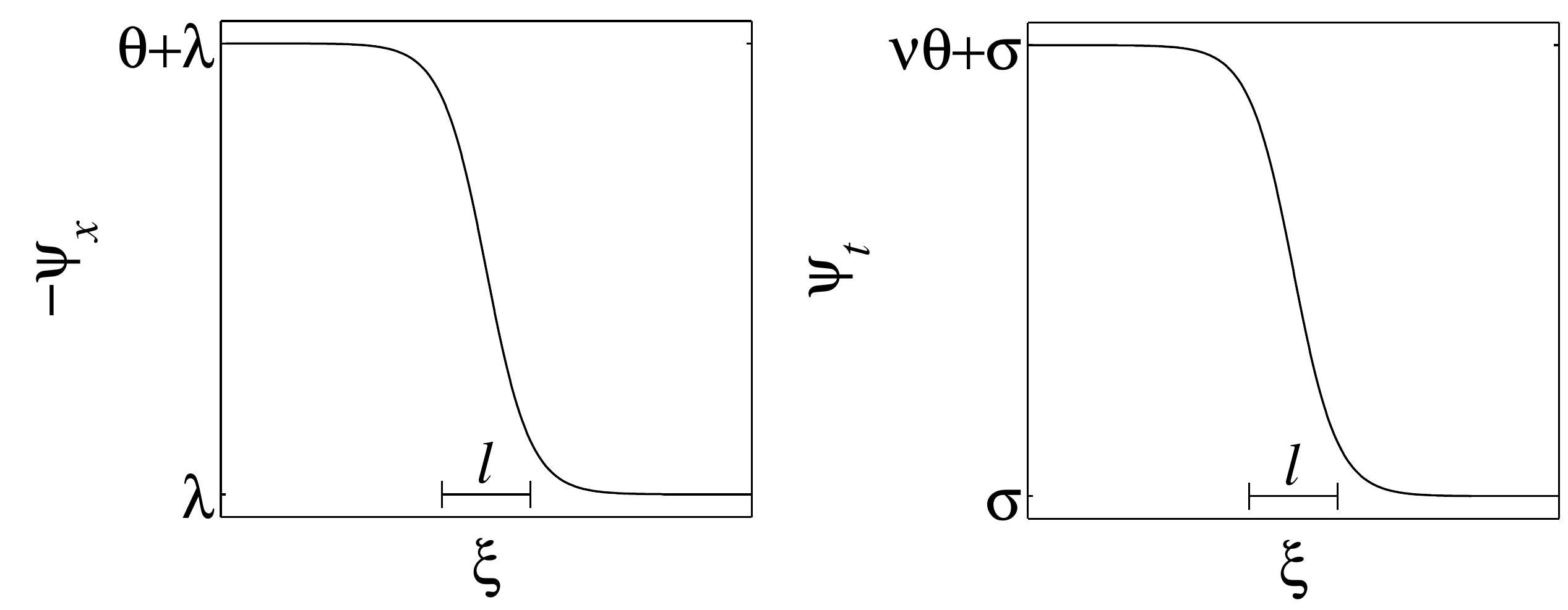}
\caption{The exact solution of Eq.~\eqref{modkuz1D} represents a
  traveling front. In order for the solution to exist, the wave propagation
  velocity, $v$, the front height, $\theta$, and the two constants, $\lambda$ and
  $\sigma$, must satisfy Eq.~\eqref{cubic}. The plot shows a front with $l>0$.
}\label{wavefront}
\end{figure}

\new{Regarding the exact solution derived above, the physical properties of the flow associated with the traveling front are obtained from the partial derivatives of Eq.~\eqref{twpsi}, which are given by}
\begin{align}\label{twpsidrv}
  -\psi_x = \Phi + \lambda \quad \text{and} \quad \psi_t = v\Phi + \sigma.
\end{align}
According to Eq.~\eqref{velpot} the fluid particle velocity is obtained as $u =
-\psi_x$, and the first order approximation of Eq.~\eqref{sop} yields the
acoustic pressure as $p-p_0 \approx \rho_0 \psi_t$. The boundary conditions of
the front are obtained from Eqs.~\eqref{twsphi} and \eqref{twpsidrv} as
\begin{subequations}\label{bcpsi}
\begin{align}
  -\psi_x &\to \left\{\begin{array}{ll} \theta + \lambda, &\xi \to \mp \infty \\
      \lambda, &\xi \to \pm \infty
\end{array}\right., \label{bcpsia} \\ 
\psi_t &\to \left\{\begin{array}{ll} v\theta+\sigma , &\xi \to \mp \infty \\
    \sigma, &\xi \to \pm \infty
\end{array}\right., \label{bcpsib}
\end{align}
\end{subequations}
where upper (lower) signs apply for $l>0$ ($l<0$). Hence, the four parameters
$v$, $\theta$, $\lambda$, and $\sigma$, that was introduced in the derivation of
the exact solution, determine the four boundary conditions of the front. \new{From these boundary conditions we find that $\theta$ and $v\theta$ correspond to the heights of the jump across the front measured in $-\psi_x$ and $\psi_t$, respectively, see Fig.~\ref{wavefront}.} At this point it is appropriate to emphasize that, in
order for the exact solution to exist, the four parameters $v$, $\theta$,
$\lambda$, and $\sigma$ must satisfy the cubic equation \eqref{cubic}.
Furthermore, the \new{allowable} values of the wave propagation velocity correspond to
the \emph{real} roots of this equation. A noticeably property of
Eq.~\eqref{cubic}, which will prove useful later on, is that the equation is
invariant under the transformation
\begin{align}\label{trf}
  v \to v, \quad \theta \to -\theta, \quad \lambda \to \theta + \lambda, \quad
  \sigma \to v\theta + \sigma.
\end{align}
Also the boundary conditions \eqref{bcpsi} are invariant, since, according to
Eq.~\eqref{thick}, the above transformation leads to $l \to -l$.

\begin{figure}[h]
  \centering \includegraphics[width=8cm]{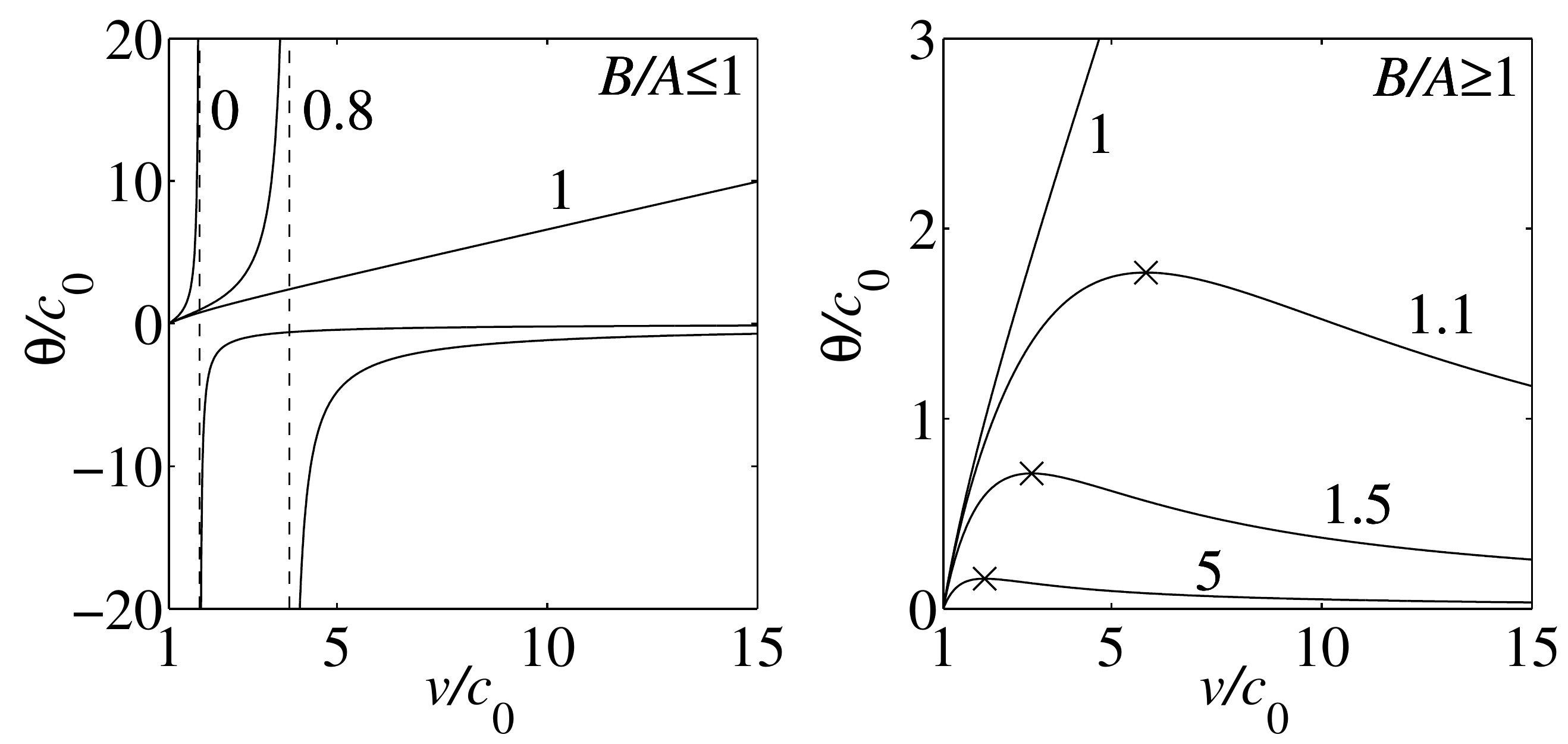}
\caption{The plots show the relationship between the front height, $\theta$, and
  the front propagation velocity, $v$, given by Eq.~\eqref{thetaOFv} with
  $\lambda=0$, $\sigma=0$, and $B/A=\{0, 0.8, 1, 1.1, 1.5, 5\}$ (see labels on
  the plots). The dashed lines indicate the singularity at $v=v_\text{s}$, which
  is defined in Eq.~\eqref{vc1}, and crosses indicate the maximum $(\theta,v) =
  (\theta_\text{max},v_\text{max})$ defined in
  Eq.~\eqref{vc2}.}\label{cubicplot}
\end{figure}

In order to investigate the relationship between the front height, $\theta$,
and the front propagation velocity, $v$, we solve Eq.~\eqref{cubic} with respect
to $\theta$ to obtain
\begin{align}\label{thetaOFv}
  \theta = \dfrac{\left(1 - \dfrac{B/A - 1}{c_0^2} \sigma \right) v^2 - 2
    \lambda v - c_0^2 - \sigma }{v \left(\dfrac{3}{2} +
      \dfrac{B/A-1}{2c_0^2}v^2\right)}.
\end{align}
For $B/A<1$ the curve $\theta(v)$ has singularities at
\begin{align}\label{vc1}
  v = v_\text{s} \equiv \pm \left(\frac{3c_0^2}{1 - B/A}\right)^{1/2},
\end{align}
and for $B/A>1$ the curve has a maximum\footnote{The critical point $(v,\theta)
  = (v_\text{max},\theta_\text{max})$ was identified by Jordan \cite{jordan1} as
  the solution bifurcation point.} at $(v,\theta) =
(v_\text{max},\theta_\text{max})$, where $v_{\text{max}}$ is obtained as
\begin{align}\label{vc2}
  v_{\text{max}} = c_0 \left(\frac{3\,B/A + \sqrt{9\,(B/A)^2 + 12(B/A - 1)}}
    {2(B/A-1)}\right)^{1/2},
\end{align}
when $\lambda=0$ and $\sigma=0$. These two \new{characteristic properties} of the curve are
illustrated in Fig.~\ref{cubicplot}.

Finally, it should be emphasized that the generalized traveling wave analysis
conducted above also applies to the one-dimensional approximation of the
Kuznetsov equation \eqref{origkuz}. In this case Eq.~\eqref{cubic} is replaced
by\footnote{Eliminating $\lambda$ and $\sigma$ from Eq.~(\ref{cuborig}) makes
  the equation equivalent to the previously reported result \cite{jordan1}}
\begin{align}\label{cuborig}
  \frac{B/A}{2c_0^2} \theta v^3 - \left(1 - \frac{B/A}{c_0^2} \sigma\right) v^2
  + \left(\theta + 2\lambda\right) v + c_0^2 = 0,
\end{align}
and Eq.~\eqref{thick} by
\begin{align}
  l = \frac{4b}{\left(\dfrac{B/A}{2c_0^2}v^2 + 1\right) \theta}.
\end{align}
\new{Apart from these changes, a generalized traveling wave analysis of the Kuznetsov equation is basically identical to that of Eq.~\eqref{modkuz1D}.}

The Hamiltonian structure, however, is unique to the proposed nonlinear wave
equation \eqref{modkuz} and its one-dimensional approximation Eq.~\eqref{modkuz1D}. In order to establish a relationship between the exact
solution, derived in this section, and the Hamiltonian structure of the
governing equation, \new{we insert Eq.~\eqref{twpsi} into Eq.~\eqref{eqnrg} and the one-dimensional
approximations of Eqs.~\eqref{hamden} and \eqref{tnrg}.} Then, after some
calculations, we find that Eq.~\eqref{eqnrg} reduces to the cubic equation
\eqref{cubic}. Hence, the exact traveling front solution of the proposed
Hamiltonian model equation \eqref{modkuz1D} satisfies the energy balance equation
\eqref{eqnrg}.

\subsection{Linear stability analysis\label{linstab}}
In order to gain insight into the stability properties of the traveling front
solution we initially consider the constant solution
\begin{align}\label{csol}
  -\psi_x = K \qquad \text{and} \qquad \psi_t = L,
\end{align}
which satisfies the nonlinear wave equation \eqref{modkuz1D}. The two constants
$K$ and $L$ are arbitrary. In order to investigate the linear stability
properties of this solution, we add small perturbation terms to the constant
values as
\begin{align}\label{stabanz}
  -\psi_x = K - \varepsilon \chi_x \qquad \text{and} \qquad \psi_t = L +
  \varepsilon \chi_t,
\end{align}
where $\chi = \chi(x,t)$ and $\varepsilon \ll 1$. Then, inserting
Eqs.~\eqref{stabanz} into Eq.~\eqref{modkuz1D} and keeping terms up to first
order in $\varepsilon$ we obtain the following linear perturbation equation
\begin{align}\label{linprtb}
  \left(1 - \frac{B/A - 1}{c_0^2} L \right) \chi_{tt} - \left(c_0^2 + L \right)
  \chi_{xx} = b \chi_{xxt} - 2 K \chi_{xt}.
\end{align}
Inserting the single Fourier mode
\begin{align}
  \chi(x,t) = D\,e^{\text{i}\left(k x - \omega t\right)},
\end{align}
where $D$ is the amplitude, $k$ is the wave number, and $\omega$ is the angular
frequency, into Eq.~\eqref{linprtb}, we obtain the following dispersion relation
\begin{multline}\label{disp}
  \left(\frac{B/A - 1}{c_0^2} L - 1\right) \omega^2 + \left(2 K k - ibk^2\right)
  \omega \\ + \left(c_0^2 + L \right) k^2 = 0.
\end{multline}
The constant solution \eqref{csol} is asymptotically stable \new{only} if all solutions of
Eq.~\eqref{linprtb} approach zero as $t \to \infty$. This is the case when the
imaginary part of both roots in Eq.~\eqref{disp}, $\omega_1$ and $\omega_2$, are
negative. It can be shown that for $B/A>1$ the only requirement in order for the
imaginary part of both roots to be negative is
\begin{align}\label{stabreg1}
-c_0^2 < L < \frac{c_0^2}{B/A-1}.
\end{align}
When $B/A<1$ the only requirement for both roots to have a negative imaginary
part is
\begin{align}\label{stabreg2}
  -c_0^2 < L < \infty.
\end{align}
Hence, the stability properties of the constant solution \eqref{csol} are
determined exclusively by $L$, i.e.\ the constant value of $\psi_t$. \new{Recall that the acoustic pressure is proportional to $\psi_t$, thus, the level of the acoustic pressure determines the stability properties of the solution.}

In order for the front solution to be stable, it is a necessary condition that
both \new{left and right} asymptotic values of $\psi_t$, given by Eq.~\eqref{bcpsib}, belong to the
interval \eqref{stabreg1} when $B/A>1$, and the interval \eqref{stabreg2} when
$B/A<1$. In Section~\ref{numstab} we shall further investigate this stability
criterion by means of numerical simulations.

\section{Front-shock relationship\label{fsr}}
Within fluid dynamics, a shock denotes a sharp change of the physical quantities. A
shock propagates at supersonic speed with respect to the fluid ahead of it,
while it remains subsonic with respect to the fluid behind it. The physical
quantities of the flow on each side of the shock are connected by the
Rankine-Hugoniot relations, which are conservation equations for mass, momentum
and energy. In the following we shall demonstrate that the front solution of the
proposed Hamiltonian model equation \eqref{modkuz1D} retains these properties.

\subsection{The Rankine-Hugoniot relations}
Using square brackets to denote the change in value of any quantity
across a shock, e.g.
\begin{align}\label{sqb}
\left[\rho\right] = \rho_\text{a} - \rho_\text{b},
\end{align}
where $\text{b}$ denotes the value behind the shock and $\text{a}$ denotes the
value ahead of it, the Rankine-Hugoniot relations may be written as \cite{landau}
\begin{align}
  \text{mass}:& \quad \left[\rho \left(u - v\right)\right] = 0,
\label{massc} \\ 
\text{momentum}:& \quad \left[p + \rho \left(u-v\right)^2\right] = 0,
\label{momc} \\ 
\text{energy}:& \quad \left[h + \left(u-v\right)^2/2 \right] = 0,
\label{enerc}
\end{align}
where $v$ is the shock propagation velocity and $h$ is the enthalpy.

We now replace $u$, $\rho$, $p$, and $h$ with expressions in terms of $\psi_x$
and $\psi_t$, and write all equations retaining terms up to second order. Upon
setting $u = -\psi_x$ and substituting Eqs.~\eqref{sorho} and \eqref{sop} into
Eqs.~\eqref{massc} and \eqref{momc} we thus obtain
\begin{multline}\label{masscpsi}
  \Biggl[\left(\frac{\left(\psi_x\right)^2}{2} +
      \frac{B/A-1}{2c_0^2}\left(\psi_t\right)^2 \right)v \\ - \psi_t \left(\psi_x +
      v\right) - c_0^2 \psi_x \Biggr] = 0,
\end{multline}
and
\begin{multline}\label{momcpsi}
  \Biggl[\frac{B/A-1}{2c_0^2} \left(\psi_t\right)^2 v^2 - \psi_t v^2 - 2 \psi_t
  \psi_x v  \\ - \frac{\left(\psi_t\right)^2}{2} - 2 c_0^2 \psi_x v - c_0^2
  \psi_t\Biggr] = 0,
\end{multline}
respectively. The dissipative terms involving $\kappa$, $\zeta$, and $\eta$ do
not appear in Eqs.~\eqref{masscpsi} and \eqref{momcpsi}, since $\psi_{xx} \to 0$
ahead of and behind the front. The changes in $\psi_x$ and $\psi_t$ across the
front are obtained from the boundary conditions \eqref{bcpsi}. Assuming that
$l>0$ and $v>0$, and using the notation introduced in Eq.~\eqref{sqb} we may
write
\begin{subequations}\label{jumpie}
\begin{align}
  \left[\psi_x\right] = \theta, \qquad \left[\psi_t\right] = -v\theta.
\end{align}
Furthermore, changes in products of $\psi_x$ and $\psi_t$ are
\begin{align}
  \left[\left(\psi_x\right)^2\right] &= -\theta^2 - 2\theta\lambda, \\
  \left[\left(\psi_t\right)^2\right] &= -v^2\theta^2 - 2v\theta\sigma, \\
  \Big[\psi_x \psi_t\Big] &= v\theta^2 + v\theta\lambda + \theta\sigma.
\end{align}
\end{subequations}
Inserting Eqs.~\eqref{jumpie} into Eqs.~\eqref{masscpsi} and \eqref{momcpsi}
\emph{both} conservation equations reduce to the cubic equation \eqref{cubic}.
This striking result leads to the conclusion, that Eq.~\eqref{cubic} implies
conservation of mass and momentum. At this point it should be noted that the
generalized traveling wave analysis of the Kuznetsov equation \eqref{origkuz}
leads to the cubic equation \eqref{cuborig}, which is \emph{not} in agreement
with the conservation equations for mass and momentum.

In order to handle the enthalpy in the condition for energy conservation
\eqref{enerc} we shall make use of the following fundamental thermodynamic
relationship \cite{landau}
\begin{align}
\nablabf h = \frac{\nablabf p}{\rho}.
\end{align}
Using the equation of motion \eqref{motion}, subject to the basic
assumptions of the derivation of the model equations in Section~\ref{nlweq}, we
obtain from Eq.~\eqref{enerc}
\begin{align}\label{tvc}
\left[\psi_t + v\psi_x\right] = 0.
\end{align}
Alternatively, Eq.~\eqref{tvc} follows directly from the generalized traveling
wave assumption \eqref{mtwas}. Hence, the traveling wave assumption implies
energy conservation in the flow.

\subsection{Sub-/supersonic speeds of propagation}
In order to determine whether the traveling front solution, derived in
Section~\ref{gtwa}, propagates at sub- or supersonic speed with respect to the
fluid ahead of it and the fluid behind it, we need to introduce the speed of
sound in these regions of the fluid. Without loss of generality, we may consider
only fronts propagating in the positive direction, $v>0$, since
Eq.~\eqref{modkuz1D} is invariant under the transformation $x \to -x$.
Furthermore, we shall limit the analysis to stable fronts, i.e.\ $\psi_t$ must
belong to the interval \eqref{stabreg1} when $B/A>1$, and the interval
\eqref{stabreg2} when $B/A<1$. Then, letting $\theta \to 0$ in Eq.~\eqref{cubic}
and solving for $v$ yields the small signal propagation velocity, which is
equivalent to the speed of sound, $c$.  Introducing $\lambda=K$ and $\sigma=L$
we obtain
\begin{align}\label{sos}
  v = c(K,L) \equiv \frac{K + \sqrt{K^2 + \left(L + c_0^2\right) \left(1 -
        \dfrac{B/A-1}{c_0^2}L\right)}} {1 - \dfrac{B/A-1}{c_0^2}L},
\end{align}
where $K$ and $L$ denote the constant levels of $-\psi_x$ and $\psi_t$, respectively, at which
the speed of sound \eqref{sos} is evaluated. Inserting the boundary conditions
of the front into Eq.~\eqref{sos}, i.e.\ substituting Eq.~\eqref{bcpsia} for $K$
and Eq.~\eqref{bcpsib} for $L$, we obtain the speed of sound ahead of,
$c_{\text{a}}$, and behind, $c_{\text{b}}$, the front
\begin{align}
  c_{_{\textstyle ^\mathrm{a}_\text{b}}} &= c(\lambda,\sigma), \label{cab1} \\
  c_{_{\textstyle ^\mathrm{b}_\text{a}}} &= c(\theta + \lambda, v\,\theta +
  \sigma),
  \label{cab2}
\end{align}
where upper (lower) subscripts apply for $l>0$ ($l<0$). Note that, under the
transformation \eqref{trf}, Eq.~\eqref{cab2} transforms into Eq.~\eqref{cab1}.
Hence, without loss of generality we shall consider only Eq.~\eqref{cab1} in the
following. 

In order to compare the front propagation velocity, $v$, to $c_{\text{a}}$ and
$c_{\text{b}}$ we make the following observations. Inserting
Eq.~\eqref{thetaOFv} into Eq.~\eqref{thick} yields
\begin{align}\label{sit}
  l = \frac{4bv}{\left(1 - \dfrac{B/A - 1}{c_0^2} \sigma \right) v^2 - 2 \lambda
    v - c_0^2 - \sigma}.
\end{align}
The denominator in Eq.~\eqref{sit} becomes zero when $v = c(\lambda,\sigma)$,
where $c(\lambda,\sigma)$ is given by Eq.~\eqref{sos}. Then, given that $v>0$,
we obtain from Eq.~\eqref{sit} that
\begin{align}\label{lvc}
  v > c(\lambda,\sigma) \Leftrightarrow l > 0 \quad \text{and} \quad v <
  c(\lambda,\sigma) \Leftrightarrow l < 0.
\end{align}
Finally, from Eqs.~\eqref{cab1} and \eqref{lvc} it follows that
\begin{align}
  v>c_\text{a} \quad \text{and} \quad v<c_\text{b}.
\end{align}
Hence, in all cases, the propagation velocity of the exact traveling front
solution is supersonic with respect to the fluid ahead of the front, and
subsonic with respect to the fluid behind it.

\section{Numerical results\label{numstud}}
All numerical calculations rely on a commercially available software
package\footnote{COMSOL version 3.2a, \texttt{http://www.comsol.com} (2005)},
which is based on the finite element method. For convenience we introduce the
following non-dimensional variables, denoted by tilde
\begin{align}\label{ndvars}
  \tilde{\psi}(\tilde{x},\tilde{t}\,) = \frac{1}{b} \psi(x,t), \qquad \tilde{x}
  = \frac{c_0}{b} x, \qquad \tilde{t} = \frac{c_0^2}{b} t.
\end{align}
Under this transformation we may write Eq.~\eqref{modkuz1D} as
\begin{multline}\label{ndmkuzeq}
  \psi_{tt} - \psi_{xx} = \psi_t \psi_{xx} \\ + \frac{\partial}{\partial t}
  \left(\psi_{xx} + \left(\psi_x\right)^2 +
    \frac{B/A-1}{2}\left(\psi_t\right)^2\right),
\end{multline}
where the tildes have been omitted. From a comparison of Eqs.~\eqref{modkuz1D} and \eqref{ndmkuzeq}, we find that the results of the previous sections
subject to $b=1$ and $c_0=1$ apply to Eq.~\eqref{ndmkuzeq}. Non-dimensional
versions of the parameters, $v$, $\theta$, $\lambda$, and $\sigma$, also
indicated by tilde, become
\begin{align}
  \tilde{\lambda} = \frac{\lambda}{c_0}, \qquad \tilde{\sigma} =
  \frac{\sigma}{c_0^2}, \qquad \tilde{\theta} = \frac{\theta}{c_0}, \qquad
  \tilde{v} = \frac{v}{c_0}.
\end{align}
In the following analysis we consider only the non-dimensional formulation of the
problem. For notational simplicity we shall omit the tildes.

\subsection{Investigation of the front stability criterion\label{numstab}}
In order to investigate, numerically, the stability properties of the front, we
chose as initial condition for the numerical solution, the exact solution given
by Eqs.~\eqref{twpsi} and \eqref{twpsidrv}, and choose $v$, $\theta$, $\lambda$,
$\sigma$, and $B/A$ such that Eq.~\eqref{cubic} is satisfied. For the sake of
clarity, we shall limit the numerical investigations to the specific case of
$\lambda=0$, $\sigma=0$, $v>0$, and $l>0$, which, according to
Eq.~\eqref{bcpsi}, corresponds to fronts propagating to the right into an
unperturbed fluid.

\begin{figure}[h]
  \centering \includegraphics[width=8cm]{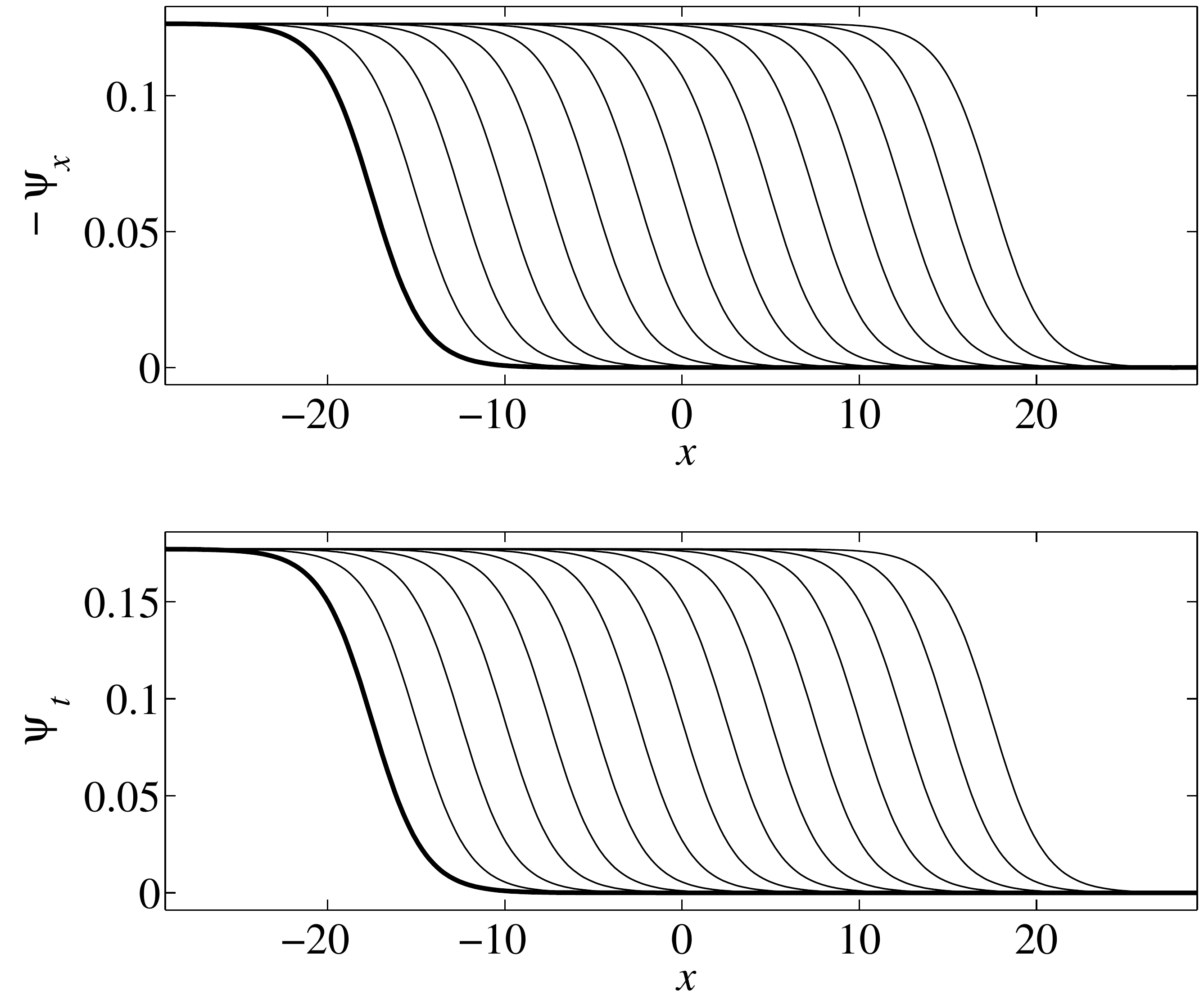}
\caption{The initial condition at $t=0$ (bold line) is
  obtained from Eqs.~\eqref{twpsi} and \eqref{twpsidrv} subject to $v=1.4$,
  $\theta=0.127$, $\lambda=0$, $\sigma=0$, and $B/A=5$, which satisfy
  Eq.~\eqref{cubic}. The numerical solutions are shown over the time interval $0
  \leq t \leq 25$.}
\label{front}
\end{figure}

\begin{figure}[h]
  \centering \includegraphics[width=8cm]{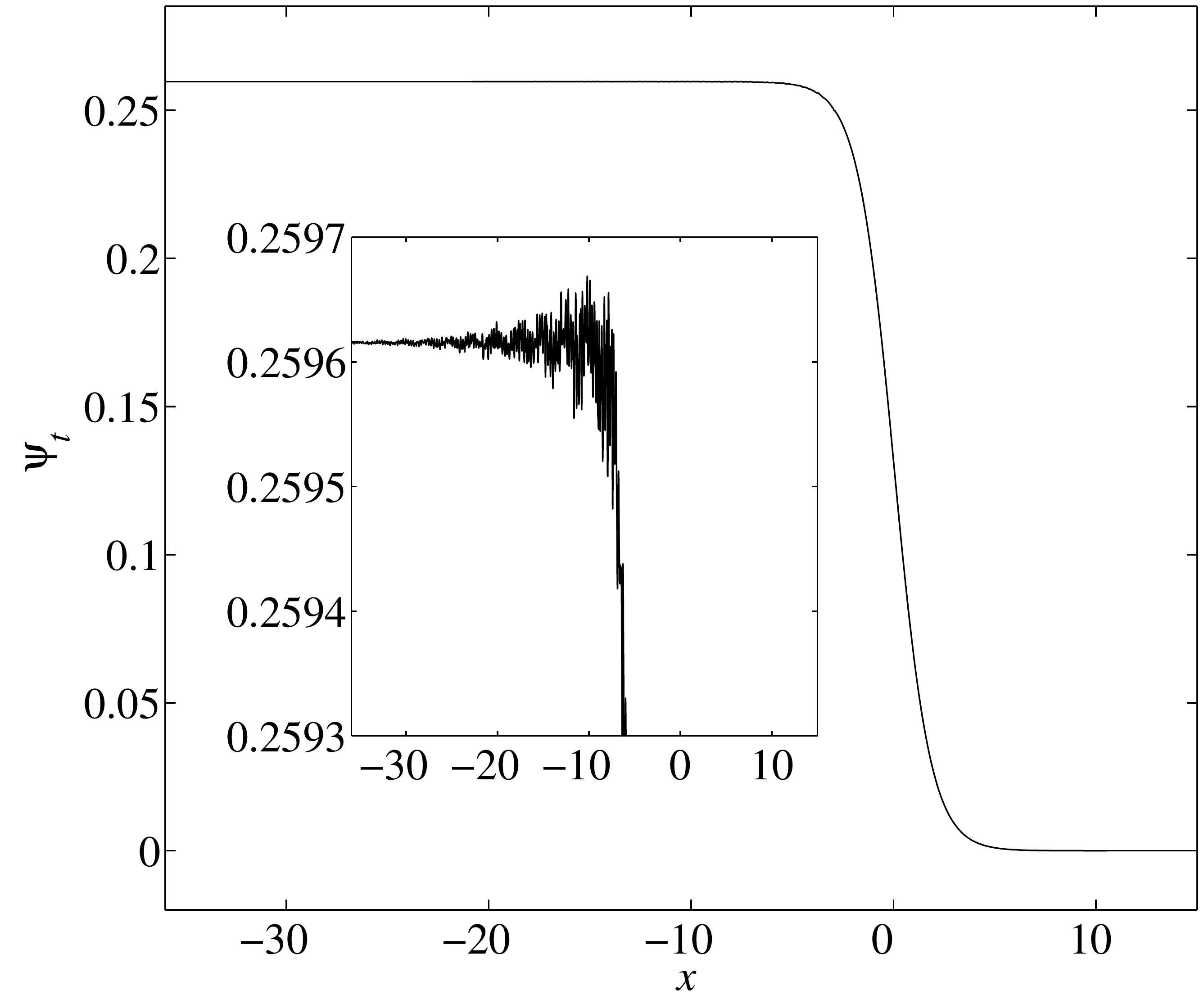}
\caption{The numerical algorithm fails to integrate this solution
  forward in time (insert shows magnification). See caption of Fig.~\ref{front}.
  $v=1.7$, $\theta=0.153$, $\lambda=0$, $\sigma=0$, and $B/A=5$. The numerical
  solution is shown at time $t = 3.8 \E{-3}$.}\label{unstabbie}
\end{figure}

A first numerical simulation is presented in Fig.~\ref{front}. Evidently, the
numerical algorithm successfully integrates the initial condition forward in
time. This finding indicates that, for the specific choice of parameters,
$v=1.4$, $\theta=0.127$, and $B/A=5$, the front exists and is stable. A second
numerical simulation is presented in Fig.~\ref{unstabbie}. This initial
condition is given a larger velocity, $v=1.7$, and a larger height,
$\theta=0.153$, while $B/A=5$ remains unchanged compared to the first example.
The parameters are chosen such that Eq.~\eqref{cubic} remains satisfied.
Apparently, the numerical algorithm fails when integrating the solution forward
in time, which indicates that the front is unstable for the specific choice of
parameters. Given that $\sigma=0$ and $l>0$, the left and right asymptotes of
the front are given by $\psi_t=v\theta$ and $\psi_t=0$, respectively, according
to Eq.~\eqref{bcpsib}. Clearly, the right value belongs to the interval
\eqref{stabreg1}, thus, it does not causes instability of the front. However, if
the left value, $v\theta$, lies outside the interval \eqref{stabreg1}, it causes
instability of the front. Inserting the value of $B/A$ from the two examples
above into Eq.~\eqref{stabreg1}, we find that in the first and second example,
$v\theta$ lies inside and outside the interval \eqref{stabreg1}, respectively.
Hence, the behavior observed in Figs.~\ref{front} and \ref{unstabbie} agrees
with the stability criterion introduced in Section~\ref{linstab}.

\begin{figure}[h]
  \centering \includegraphics[width=8cm]{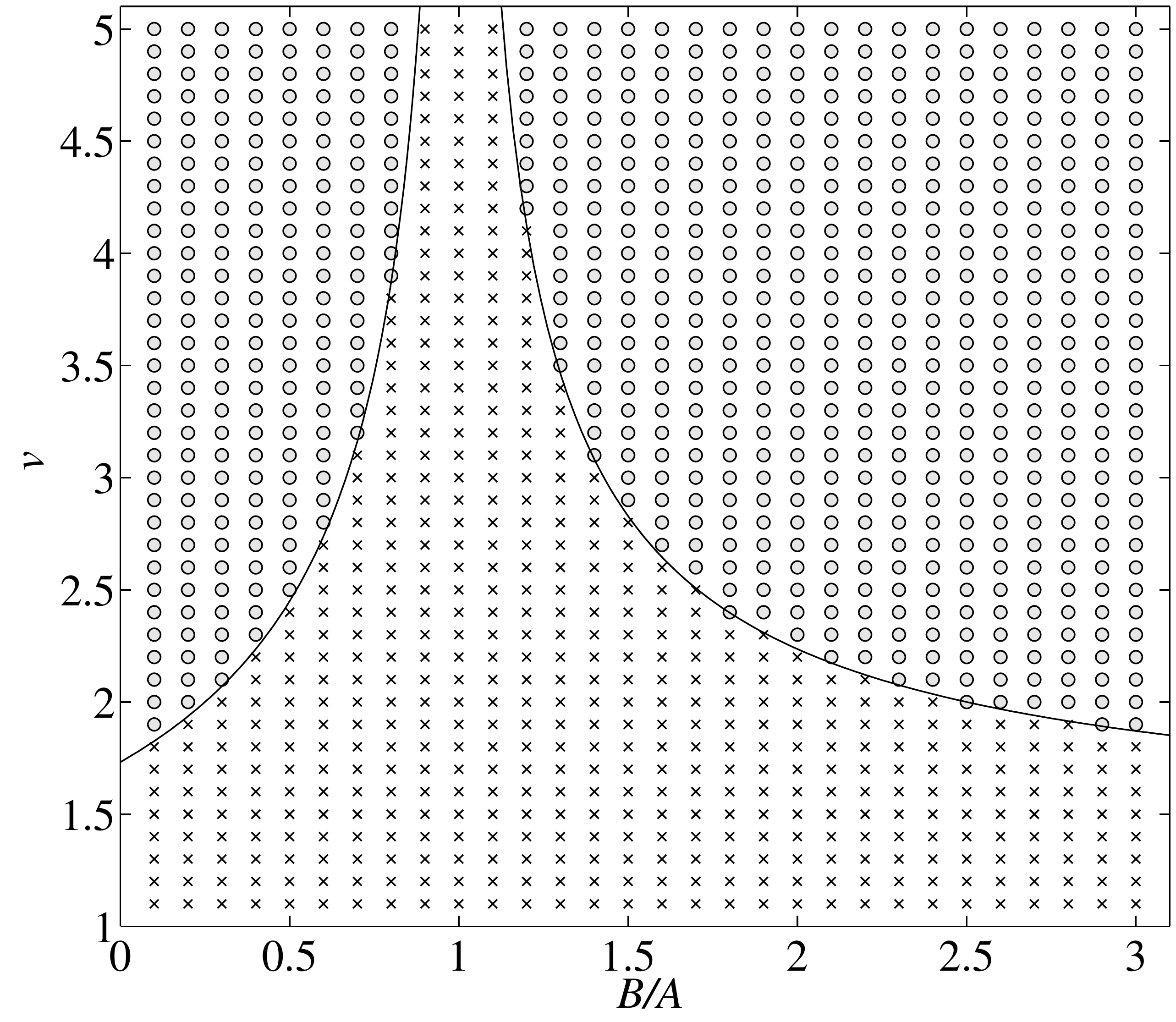}
\caption{Each point in the $(B/A,v)$-plane represents a numerical simulation,
  with the initial condition obtained from Eqs.~\eqref{twpsi} and
  \eqref{twpsidrv} subject to $\lambda=0$, $\sigma=0$, and $\theta$ given by
  Eq.~\eqref{thetaOFv}. Crosses and circles indicate stable and unstable
  solutions, respectively (compare with Figs.~\ref{front} and \ref{unstabbie}).
  Solid lines represent the stability threshold curves given by Eqs.~\eqref{vc1}
  and \eqref{vc3}.}\label{stabr}
\end{figure}

A large number of numerical simulations have been performed in order to
systematically investigate the stability properties of the front. Within each
simulation the parameters in the initial condition are, again, chosen such that
Eq.~\eqref{cubic} is satisfied. The result of this investigation is presented in
Fig.~\ref{stabr}. Still, $\sigma=0$ and $l>0$, such that the left asymptotic
value of the front is given by $\psi_t=v\theta$. For $B/A>1$ the stability
threshold curve in the $(B/A,v)$-plane is obtained when $v\theta$ equals the
upper bound of the interval \eqref{stabreg1}. Using Eq.~\eqref{thetaOFv} we
obtain
\begin{align}\label{vc3}
  v\theta = \frac{1}{B/A-1} \Rightarrow v = \frac{\sqrt{(B/A-1)(2\,B/A +
      1)}}{B/A-1}.
\end{align}
For $B/A<1$, the stability threshold curve is given by Eq.~\eqref{vc1}, since
$v\theta$ lies within (outside) the interval \eqref{stabreg2} when
$v<v_\text{s}$ ($v>v_\text{s}$), according to Eq.~\eqref{thetaOFv}. The two
stability threshold curves are included in Fig.~\ref{stabr}. The fine agreement
between the numerical results and the stability threshold curves indicates that
the stability criterion, introduced in Section~\ref{linstab}, is both necessary
\emph{and} sufficient in order for the front solution to be stable.

\subsection{Head-on colliding fronts\label{headsec}}
\begin{figure}[h]
  \centering
  \includegraphics[width=8cm]{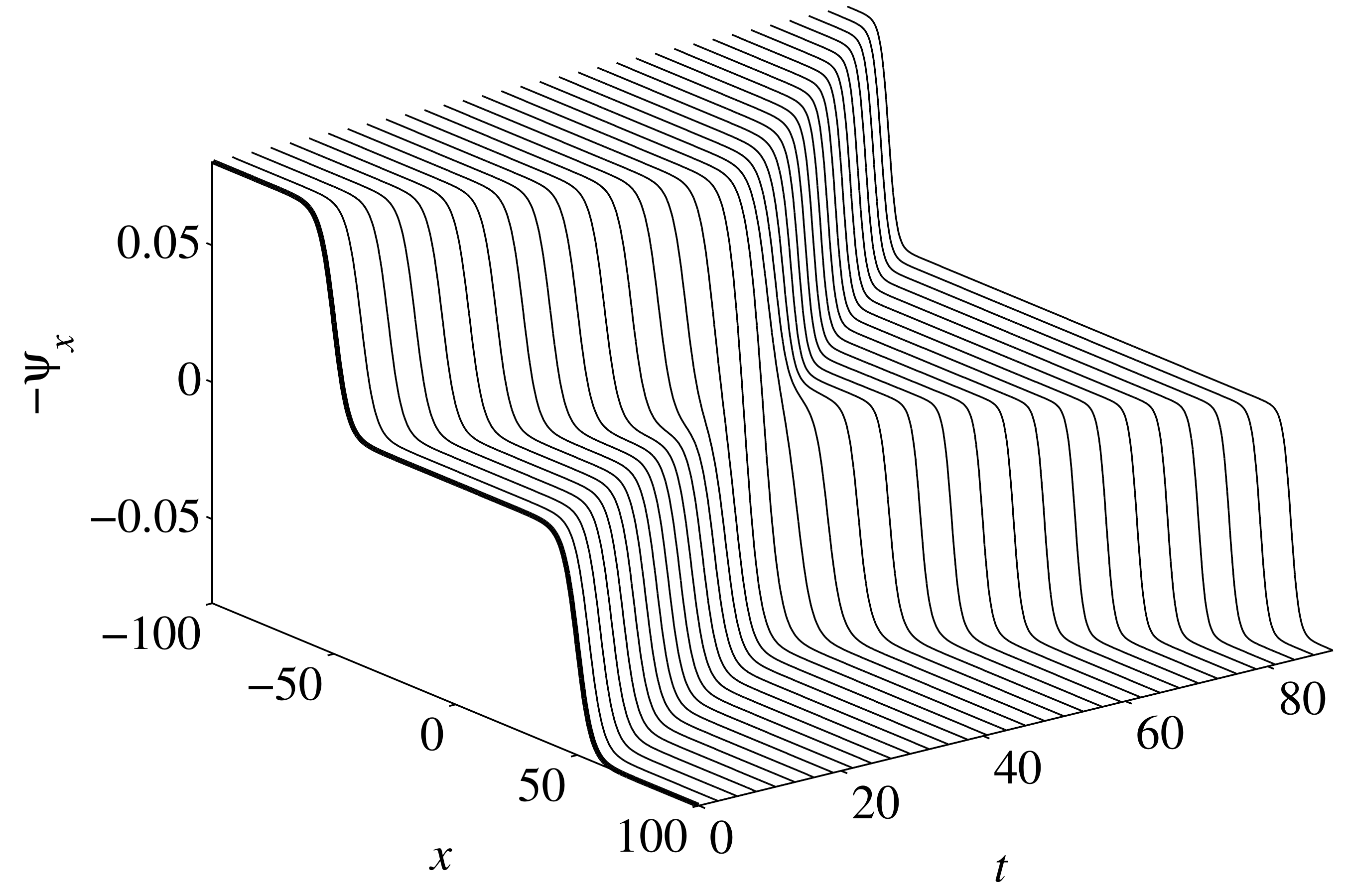}\\
  \includegraphics[width=8cm]{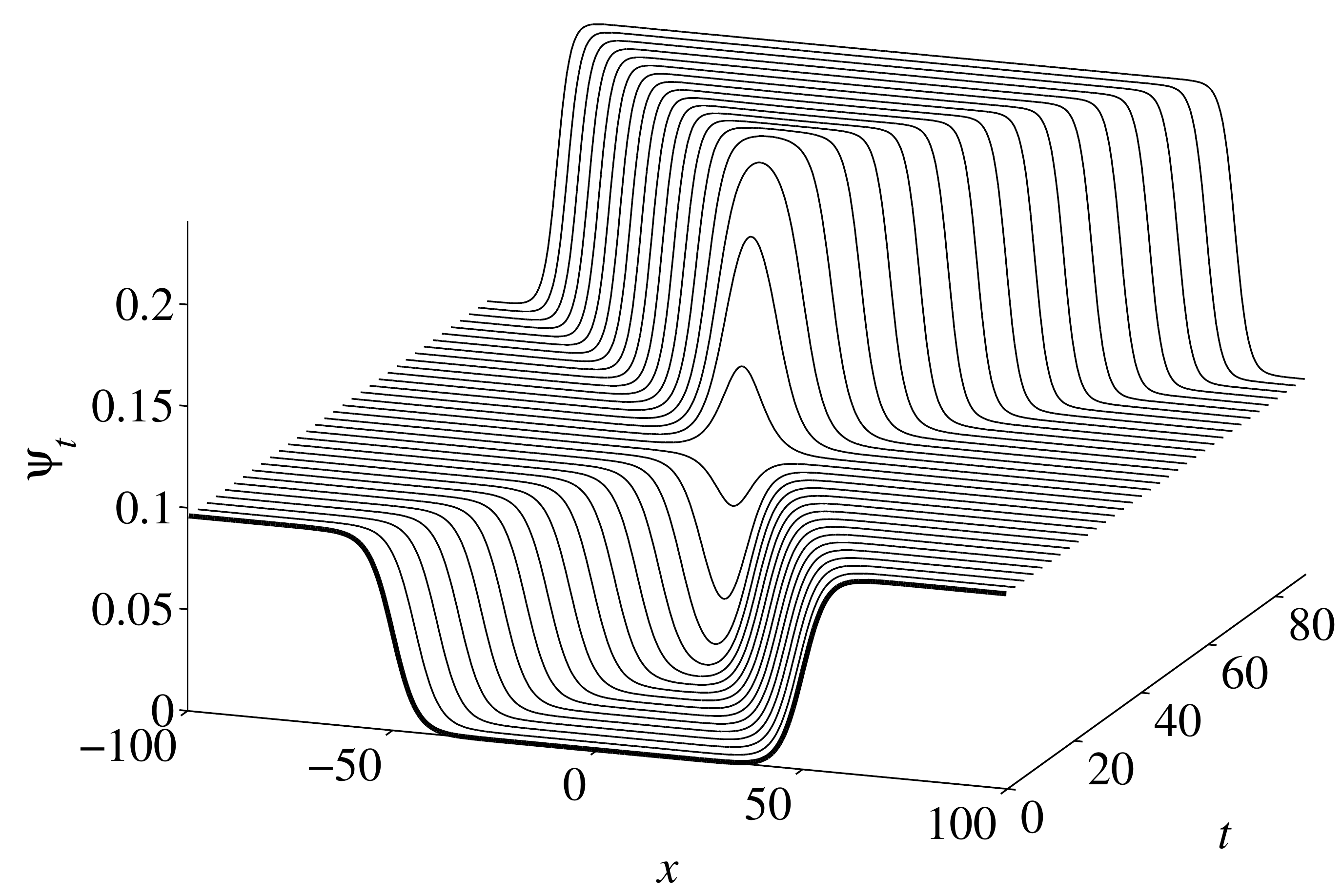}\\
  \includegraphics[width=8cm]{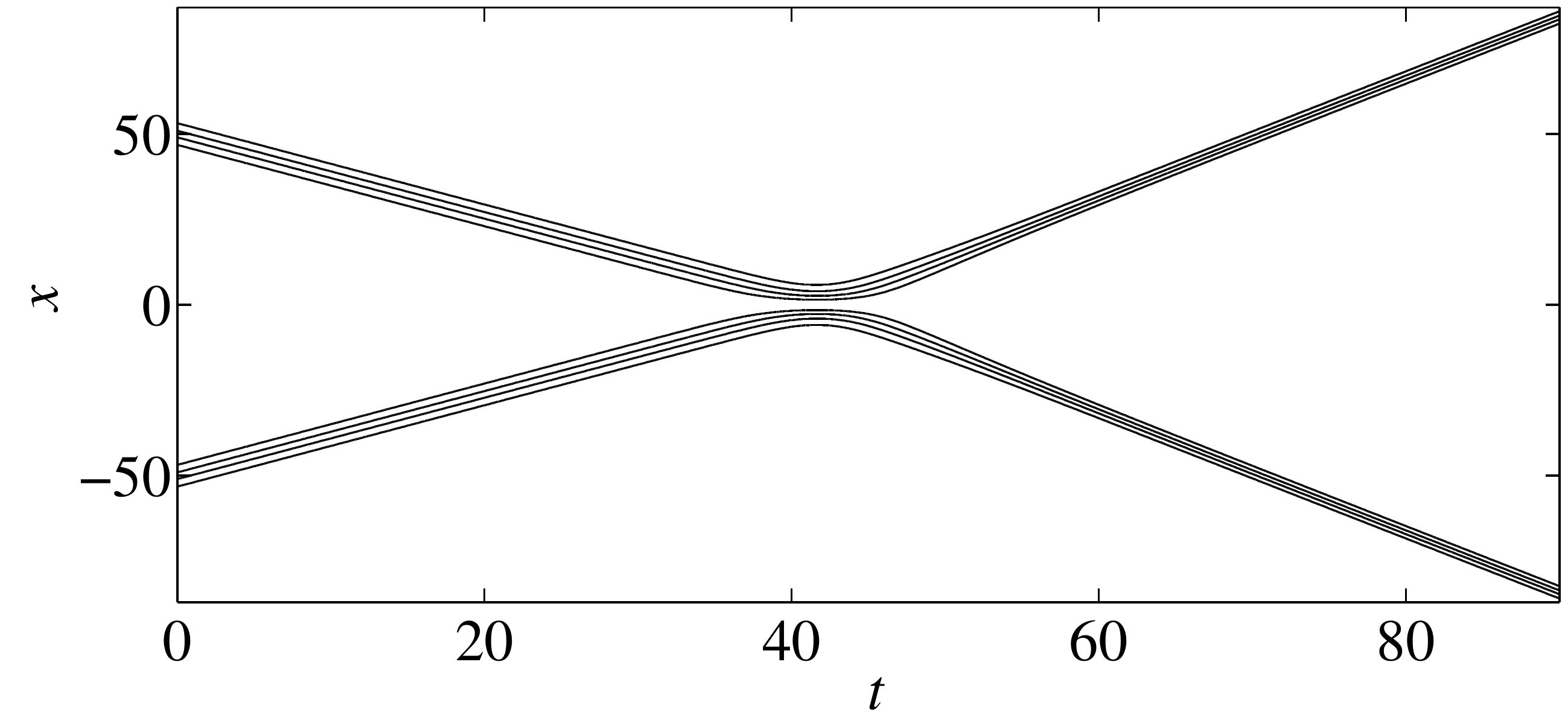}
\caption{The initial condition (bold lines in the two topmost plots),
  corresponds to two fronts that make a head-on collision at $t=42$. The initial
  fronts are defined by $v_1 = -v_2 = 1.19$, $\theta_1 = -\theta_2=8.07\E{-2}$,
  $\lambda_1 = \lambda_2 = 0$, $\sigma_1 = \sigma_2 = 0$, and $B/A=5$, where
  subscript 1 and 2 relate to fronts positioned the left and right,
  respectively. For each of the two fronts the parameters satisfy
  Eq.~\eqref{cubic}. Lowermost: contour lines given by $-\psi_x=Z$, where $Z$
  takes 4 equidistantly spaced values across each front.}
\label{headOS}
\end{figure}

The numerical simulation presented in Fig.~\ref{headOS} shows the result of a
head-on collision between two fronts. From the simulation we observe that two
new fronts emerge upon the collision. The contour plot reveals that these fronts
travel at a higher speed, compared to the speed of the fronts before the
collision. For other choices of initial condition, we found the outcome of the
head-on collision to be fronts traveling at lower speed, compared to that of the
fronts before the collision.

In order to analyze solutions of Eq.~\eqref{modkuz1D} that comprise two fronts,
we assume that these fronts belong to the class of exact front solutions derived
in Section~\ref{gtwa} above. Investigations of the fronts that emerge upon a
head-on collision have made it clear that this assumption is true, only when the
generalized traveling wave assumption is considered, in contrast to the standard
traveling wave assumption. Then, for each of the two fronts in the solution we
introduce four new parameters, $v$, $\theta$, $\lambda$, and $\sigma$, which
must satisfy Eq.~\eqref{cubic} as
\begin{multline}\label{cubics}
  \frac{B/A-1}{2} \theta_i v_i^3 - \left(1 - (B/A-1) \sigma_i\right) v_i^2 \\ +
  \left(\frac{3}{2} \theta_i + 2\lambda_i\right) v_i + \sigma_i + 1 = 0, \qquad
  i=1,2,
\end{multline}
where subscript 1 and 2 denote parameters associated with waves positioned to
the left and right, respectively. Furthermore, we require that solutions
comprising two fronts are continuous and satisfy the following set of arbitrary
boundary conditions
\begin{align}
  -\psi_x \to \left\{\begin{array}{ll} P, &x \to - \infty \\
      Q, &x \to + \infty
\end{array}\right., \quad
\psi_t \to \left\{\begin{array}{ll} R , &x \to - \infty \\
    S, &x \to + \infty
\end{array}\right..
\end{align}
Assuming that $l_1>0$ and $l_2<0$, we find, using Eq.~\eqref{bcpsi}, that the
these requirements lead to the following conditions
\begin{subequations}\label{twoCond}
\begin{gather}
  \lambda_1 = \lambda_2 = P - \theta_1 = Q - \theta_2, \\
  \sigma_1 = \sigma_2 = R - v_1\theta_1 = S - v_2\theta_2, \\
  \theta_1 + \theta_2 = P - Q, \quad v_1\theta_1 + v_2\theta_2 = R - S.
\end{gather}
\end{subequations}
Then, we substitute the boundary values found in Fig.~\ref{headOS} for $P$, $Q$,
$R$, and $S$ in Eqs.~\eqref{twoCond}, and substitute the value of $B/A$ into
Eq.~\eqref{cubics}. Finally, solving the system of equations \eqref{cubics} and
\eqref{twoCond}, we obtain the results listed in Table~\ref{tabtwo}. The
solution in the first row of the table corresponds to the initial fronts found
in Fig.~\ref{headOS}. The solution in the second row corresponds to two unstable
fronts, according the stability criterion discussed above. The two fronts that
emerge upon the head-on collision are defined by the values found in the third
row of the table. Hence, the fronts after the collision travel at the velocities
$-v_1=v_2=1.76$, which is in agreement with the velocities determined from the
slope of the contour lines in Fig.~\ref{headOS}.

\begin{table*}
\caption{Solution of Eqs.~\eqref{cubics} and \eqref{twoCond} subject to 
$P=-Q=8.07\E{-2}$, $R=S=9.60\E{-2}$, and $B/A=5$ (compare with Fig.~\ref{headOS}).} 
\label{tabtwo}
\centering
\begin{ruledtabular}
\begin{tabular}{c k c k c c c}
Solution & 
\multicolumn{1}{c}{$v_1$} & 
\multicolumn{1}{c}{$\theta_1$} & 
\multicolumn{1}{c}{$v_2$} &
\multicolumn{1}{c}{$\theta_2$} &
\multicolumn{1}{c}{$\lambda_1=\lambda_2$} & 
\multicolumn{1}{c}{$\sigma_1=\sigma_2$}  \\ \hline
1 &  1.19 & 8.07\e{-2} & -1.19 & -8.07\e{-2} & 0 & 0 \\
2 & -3.25 & 8.07\e{-2} &  3.25 & -8.07\e{-2} & 0 & 35.8\e{-2} \\
3 & -1.76 & 8.07\e{-2} &  1.76 & -8.07\e{-2} & 0 & 23.8\e{-2}
\end{tabular}
\end{ruledtabular}
\end{table*}

\section{Conclusions\label{concl}}
A nonlinear wave equation that governs finite amplitude acoustic disturbances in
a thermoviscous Newtonian fluid, and includes nonlinear terms up to second
order, has been presented. The single dependent variable is the velocity
potential. It has been demonstrated that, in the non-dissipative limit, the
equation preserves the Hamiltonian structure of the fundamental fluid dynamical
equations, hence, the model equation is associated with corresponding Lagrangian
and Hamiltonian densities. Furthermore, we found that the Kuznetsov equation is
an approximation of the proposed nonlinear wave equation. However, in the
non-dissipative limit the Kuznetsov equation is not Hamiltonian. Exact traveling
front solutions, for the partial derivatives with respect to space and time of
the dependent variable, has been obtained using a generalized traveling wave
assumption. This generalized assumption leads to a wider class of exact
solutions compared the one obtained from a standard traveling wave assumption,
since the generalized assumption includes two arbitrary constants, which are
added to the partial derivatives. As a result of the generalized traveling wave
analysis we found that, in order for the front to exist, its boundary values,
its propagation velocity, and the physical parameters of the problem must
satisfy a given cubic equation in the front propagation velocity. The derivation
of the exact solution applies equally well to the proposed Hamiltonian model
equation and the Kuznetsov equation.  Results for both equations have been
given.

It has been demonstrated that the overall stability properties of the front are
determined by the stability of the two asymptotic tails of the front. A linear
stability analysis of these steady parts of the solution revealed that the front
is stable when the partial derivative with respect to time, which is
proportional to the acoustic pressure, belongs to a critical interval, and is
otherwise unstable. This stability criterion has been verified numerically, by
using the exact front solution as initial condition in a number of numerical
simulations.

It has been demonstrated that, in all cases, the front propagates at supersonic
speed with respect to the fluid ahead of it, while it remains subsonic with
respect to the fluid behind it. The same properties have been reported for the
classical fluid dynamical shock. Furthermore, it has been demonstrated that the
cubic equation, mentioned above, is equivalent to the well established
Rankine-Hugoniot relations, which connect the physical quantities on each side
of a shock.  However, this result was accomplished only when considering the
cubic equation obtained from the analysis of the proposed Hamiltonian wave
equation. The generalized traveling wave analysis based on the Kuznetsov
equation is not in agreement with the Rankine-Hugoniot relations. Estimates of
the front thickness may be obtained using the values for the diffusivity of
sound listed in Table.~\ref{parval}. In water and air front thicknesses are
found to be of the order $10^{-9}$ and $10^{-7}$ meters, respectively. However,
caution should be taken with these estimates, as the small length scales
violates the continuum assumption of the governing equations.

Numerical simulations of two head-on colliding fronts have demonstrated that two
new fronts emerge upon the collision, and that these fronts, in the general
case, travel at speeds, which are different from the speeds of the fronts before
the collision. It has been demonstrated that the velocities of the fronts after
the collision may be calculated, based on information about the fronts before
the collision. However, in order to
accomplish this calculation, it has proven necessary to introduce the
generalized traveling wave assumption in the derivation of the front solution.

In future studies, it would be rewarding to further investigate a variety of
interacting fronts, other than the head-on collision reported in this paper.
Also a search for other types of wave solutions, might learn us more about the
properties of the proposed Hamiltonian model equation and the Kuznetsov
equation.

\begin{acknowledgments}
  One of the authors (YuBG) would like to thank the MIDIT Center and
  Civilingeni\o r Frederik Leth Christiansens Almennyttige Fond for financial
  support.
\end{acknowledgments}

\bibliography{bib}
\bibliographystyle{jasanum}
\end{document}